\documentclass{iau}
\usepackage{graphicx}

\title[IAUS294.~~Solar-cycle precursors and predictions] %% give here short title %%
{Solar-cycle precursors and predictions}

\author[Jie Jiang]   %% give here short author list %%
{Jie Jiang}

\affiliation{Key Laboratory of Solar Activity, National Astronomical
Observatories, \\Chinese Academy of Sciences, Beijing 100012, China
\\email: {\tt jiejiang@nao.cas.cn}}

\pubyear{2012}
\volume{294}  %% insert here IAU Symposium No.
\pagerange{119--126}
\jname{Title of your IAU Symposium}
\editors{A.C. Editor, B.D. Editor \& C.E. Editor, eds.}
\begin{document}

\maketitle

\begin{abstract}
The sunspot number data during the past 400 years indicates that
both the profile and the amplitude of the solar cycle have large
variations. Some precursors of the solar cycle were identified
aiming to predict the solar cycle. The polar field and the
geomagnetic index are two precursors which are received the most
attention. The geomagnetic variations during the solar minima are
potentially caused by the solar polar field by the connection of the
solar open flux. The robust prediction skill of the polar field
indicates that the memory of the dynamo process is less than 11 yrs
within the framework of the Babcock-Leighton flux transport dynamo.
One possible reason to get the short magnetic memory is the high
magnetic diffusivity in the convective zone. Our recent studies show
that the radial downward pumping is another possible reason. Based
upon the mechanism, we well simulate the cycle irregularities during
RGO time period. This opens the possibility to set up a standard
dynamo based model to predict the solar cycle. In the end, the no
correlation between the polar field and the preceding cycle strength
due to the nonlinearities and random mechanisms involved in the
sunspot emergences are stressed.

%% add here a maximum of 10 keywords, to be taken form the file <Keywords.txt>
\end{abstract}

\firstsection % if your document starts with a section,
              % remove some space above using this command.
\section{Introduction}

Predicting the sunspot number began when a cyclical behavior was
noticed by Schwabe. Long-term time series observations indicate that
solar cycle amplitudes vary from cycle to cycle. This makes the
solar cycle prediction appealing. Solar cycle predictions became
important when we began putting assets in space which are directly
affected by solar activities. Further motivation to predict solar
cycle arises from the potential implications for understanding the
origin of solar magnetic field. A reliable prediction method could
provide a constraint on the dynamo mechanisms.

The paper is organized as follows. An introduction about the
solar-cycle precursors with emphases on the geomagnetic variations
and the polar field will be presented in Section 2. The dynamo
mechanisms constrained by the polar field prediction skill will be
given in Section 3. It is mainly about the flux transport mechanisms
from the poloidal field generation layer to the toroidal field
generation layer. Section 4 would be on the solar polar field
generation which is related to the generation of the poloidal field
from the toroidal field.

\section{Solar-cycle precursors}
A number of techniques are used to predict the amplitude of a cycle
during the time near or before sunspot minimum. Precursor methods
have been proved to be the most successful technique for solar
activity predictions in the past. Precursor methods are based on the
correlations between certain measured quantities in the declining
phase of a cycle and the strength of the next cycle. The geomagnetic
variation and the polar field are two precursors which are received
the most attention due to their significant prediction skill.

\subsection{Geomagnetic variation precursor}

Geomagnetic variation precursor is based on the changes in the
Earth's magnetic field. Geomagnetic indices are usually used to
quantify the geomagnetic variations. The aa-index is the simplest
three hourly global geomagnetic activity index (Mayaud, 1972). It is
estimated on the data of two geomagnetic observatories,
approximately antipodal. The main advantage in using aa-index is
that its time series is the longest available among all the
planetary indices, the first values dates back to 1868.

\begin{figure}[htb]
% \vspace*{-1.0 cm}
\begin{center}
\includegraphics[width=2.5in]{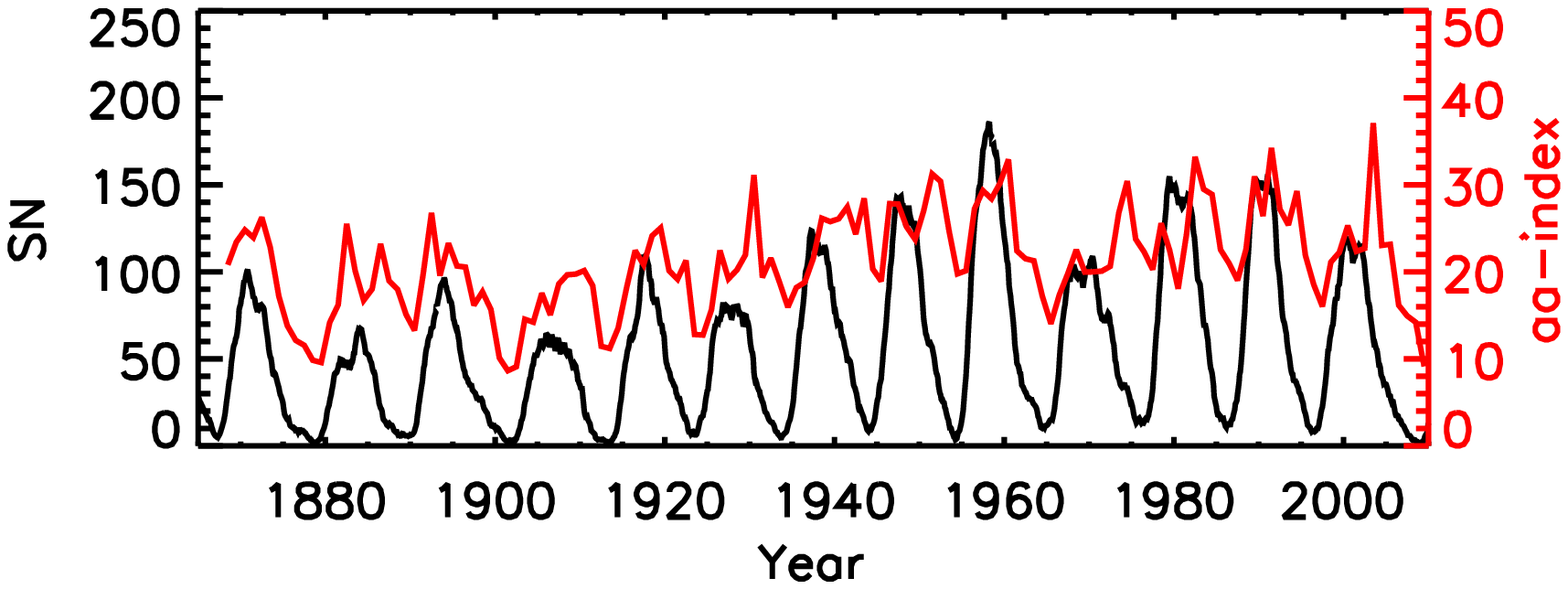}
\includegraphics[width=2.5in]{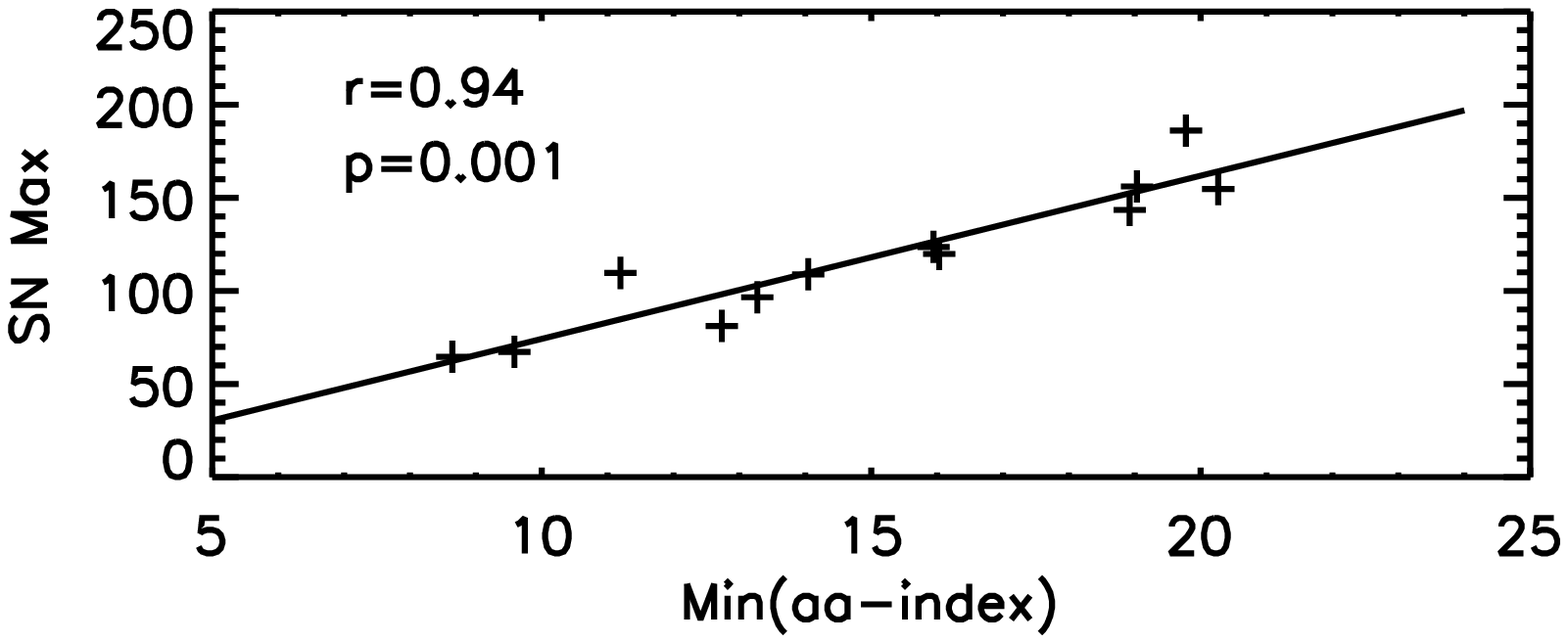}
\caption{Geomagnetic variation precursor. Left panel: time evolution
of the 12-month smoothed group sunspot number (black curve) and the
yearly aa-index (red curve). Right panel: aa-index minimum of each
cycle vs. maximum sunspot number of the subsequent cycle. The good
correlation between them provides the based of the geomagnetic
variation precursor.}
   \label{fig1}
\end{center}
\end{figure}

The left panel of Figure \ref{fig1} shows the time evolution of the
yearly aa-index data \footnote{http://www.leif.org/research/} in red
curve overplotted with 12-month smoothed group sunspot number in
black curve since 1868 onwards. The right panel shows the
correlation between the minimum values of aa-index of each cycle
with the subsequent sunspot number maximum. We may see the minimum
of the aa-index is directly related to the maximum sunspot number
for the following cycle. This provides the basis of geomagnetic
variations precursor which was firstly proposed by \cite{Ohl66}.

\subsection{Physical reasons for the geomagnetic variations}

\begin{figure}[htb]
% \vspace*{-1.0 cm}
\begin{center}
\includegraphics[width=2.5in]{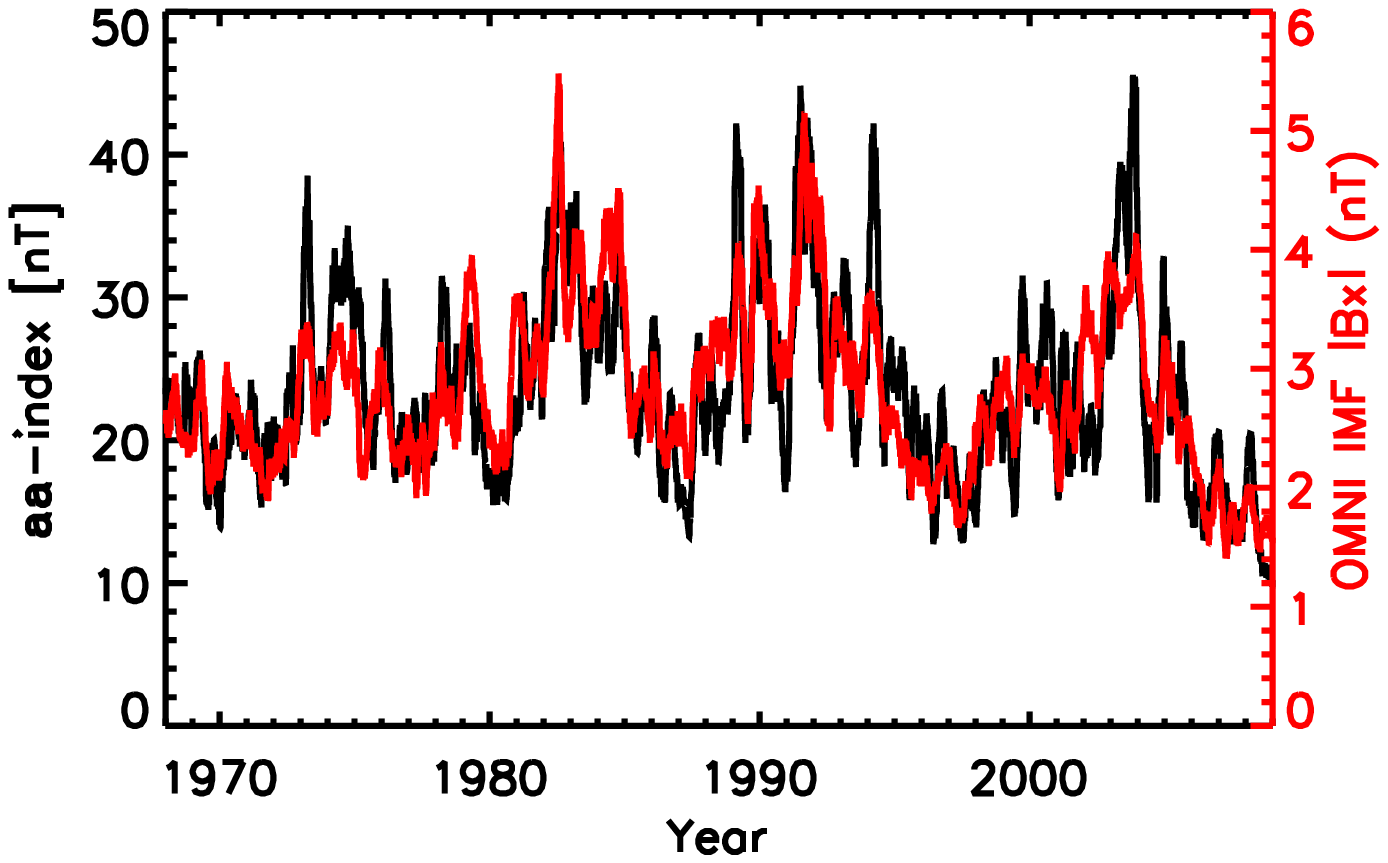}
\includegraphics[width=2.5in]{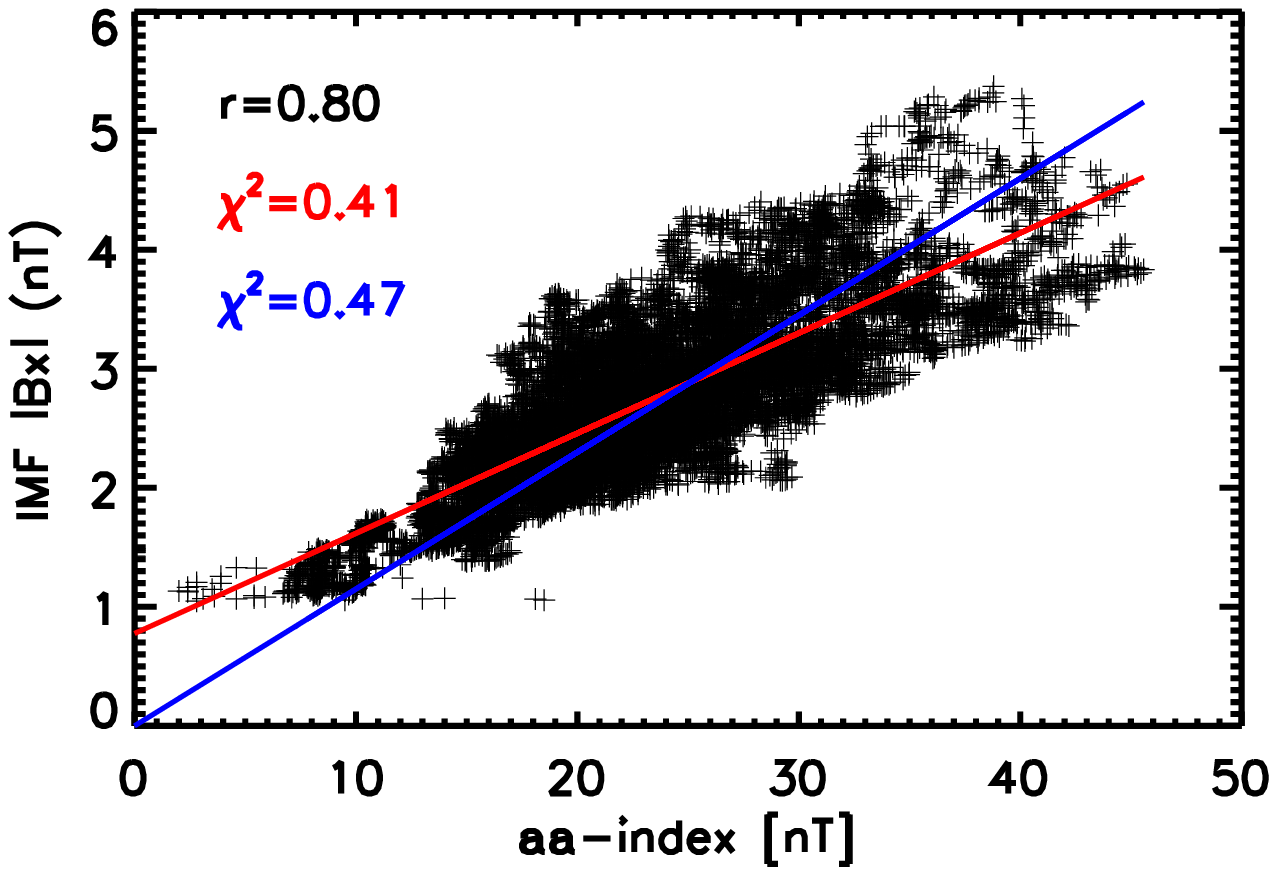}
% \vspace*{-0.5 cm}
 \caption{aa-index vs. radial component of the IMF.
Left panel: time evolution of the daily averaged aa-index (black
curve) overplotted with the daily averaged radial component of the
IMF measured by OMNI (red curve) during 1968-2010. Right panel:
correlation diagram of the aa-index and the radial IMF. The
coefficient reaches 0.8.}
   \label{fig2}
\end{center}
\end{figure}

Now let's see what causes the variation of aa-index. Left panel of
Figure \ref{fig2} shows the time evolution of the daily-averaged
aa-index overplotted with the daily-averaged radial component of the
interplanetary magnetic field (IMF) measured by OMNI
\footnote{http://omniweb.gsfc.nasa.gov/}. We may see that they
closely follow each other. Right panel shows their correlation, the
coefficient of which reaches 0.8. Hence, the aa-index variations are
almost entirely due to the variations in the near-Earth radial IMF.
This is consistent with the result given by \cite{Stamper99} who
analyzed the cause of the century-long increase in the aa-index
using observations of interplanetary space, galactic cosmic rays,
the ionosphere and the auroral electrojet.

\begin{figure}[htb]
% \vspace*{-1.0 cm}
\begin{center}
\includegraphics[width=2.5in]{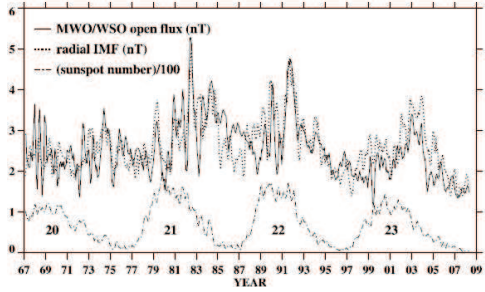}
% \vspace*{-0.5 cm}
 \caption{Comparison between the near-Earth radial IMF strength (dotted curve)
 and the total open flux, derived from a PFSS extrapolation of MWO and WSO
magnetograph measurements and expressed as an equivalent field
strength (nT) at 1 AU (solid curve). Also plotted is the sunspot
number in solid and dotted curve (from \cite[Wang 2009]{Wang09a}). }
   \label{fig3}
\end{center}
\end{figure}

The IMF is produced by the solar wind which drags some magnetic
flux, corresponding to the solar open flux, out of the Sun to fill
the heliosphere. The solid curve in Figure \ref{fig3} is the solar
open field derived by extrapolation of Mount Wilson (MWO) and Wilcox
solar (WSO) observatories magnetogram and the overplotted dashed
curve is the radial component of the IMF at 1AU (\cite[Wang \&
Sheeley 2009]{Wang09}). They show reasonably good agreement both in
its magnitudes and in the shape of its fluctuations. Hence the
variation of the radial IMF is caused by the solar open field.

Now let us analysis the solar open field. The radial magnetic field
outside of the solar surface is in the following form under the
assumption of the potential field,
\begin{equation}
B_r(r,\theta,\phi)=\sum_{l=0}^{\infty}\sum_{m=-l}^{m=l}
c_l(r)a_{lm}Y_{lm}(\theta,\phi).
\end{equation}
 The coefficient $c_l(r)\propto r^{-(l+2)}$ rapidly falls with $l$.
Hence only the lowest-order multipoles contribute significantly to
the field at the source surface and hence to the solar open flux.
The lowest-order multipoles are the axial dipole field and the
equatorial dipole field.

\begin{figure}[htb]
% \vspace*{-1.0 cm}
\begin{center}
\includegraphics[width=1.5in,angle=90]{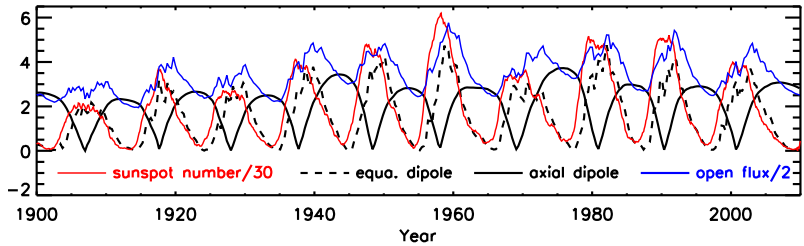}
% \vspace*{-0.5 cm}
 \caption{Reconstructed solar open flux (blue curve), equatorial dipole field
 (black dashed curve) and axial dipole field (black solid curve) since 1900 onward
 to show the relation between the equatorial and dipole field with the solar cycle.
 Also plotted is the sunspot number in red curve (adapted from Jiang et al., 2011b
 using the semi-synthetic records of emerging sunspot groups by Jiang et al., 2011a)}
   \label{fig4}
\end{center}
\end{figure}

Red curve and blue curve in Figure \ref{fig4} correspond to the time
evolution of sunspot number and the open flux during 1900-2010,
respectively. Black solid curve is the evolution of the axial dipole
field and black dashed curve is the evolution of the equatorial
dipole field. They are adapted from \cite{Jiang11b} who physically
reconstructed the solar total, polar and total flux since 1700
onward based on the semi-synthetic records of emerging sunspot
groups (\cite[Jiang et al. 2011a]{Jiang11a}). We may see that the
axial dipole is roughly anti-phase with the solar cycle which is
similar to the polar field behavior and provides the most open flux
near the cycle minimum. The equatorial dipole field is roughly in
phase with the solar cycle and provides the most open flux near the
cycle maximum.

Here is a summary about the physical reason for the geomagnetic
(aa-index) variations. Solar magnetic field varies with the solar
cycle. Solar wind drags some solar open magnetic field out of the
Sun to form the IMF. The variation of the radial IMF causes the
variation of geomagnetic (aa-) index which presents different
behaviors during different phases of solar cycle. At the cycle
maximum, aa-index reflects the strength of the solar cycle, which is
dominated by the solar equatorial dipole field. At the cycle
minimum, aa-index reflects the strength of the polar field, which is
dominated by the solar axial dipole field. Hence the behavior of the
aa-index around the solar minima is consistent with the polar field
around the solar minimum. In fact, the minimum values of aa-index
always 1-2 yr lag behind the epoch of the solar minimum (Du et al.,
2009). This is mainly due to the decay of the equatorial dipole
field on a timescale of ~1yr, see Wang \& Sheeley (2002, 2009) for
details.

\subsection{Solar polar field precursor}

\begin{figure}[htb]
% \vspace*{-1.0 cm}
\begin{center}
\includegraphics[width=2.in]{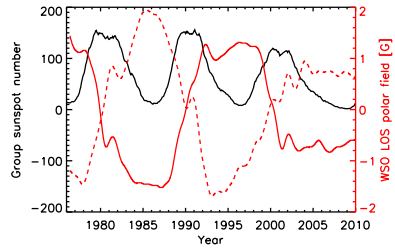}
% \vspace*{-0.5 cm}
 \caption{Time evolution of the line-of-sight solar polar field observed by WSO
 since 1976 (south pole: red dashed curve,
 north pole: red solid curve). Sunspot number is overplotted in black curve.}
   \label{fig5}
\end{center}
\end{figure}

\begin{figure}[htb]
% \vspace*{-1.0 cm}
\begin{center}
\includegraphics[width=4.in]{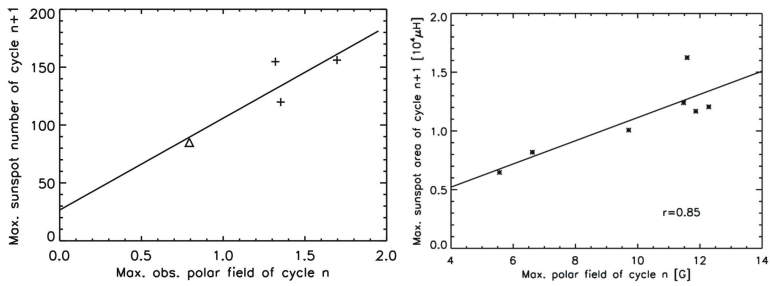}
% \vspace*{-0.5 cm}
 \caption{Correlations between the maximum polar field near the end of solar cycle $n$
 and the maximum strength of cycle $n+1$. Left: observed polar field
 during cycles 20-23 (LOS component, averaged over 35$^\circ$ wide polar caps).  Right:
 reconstructed polar field during cycles 15-21 (radial component, averaged over the 15$^\circ$ wide polar
caps, from Cameron et al., 2010).}
   \label{fig6}
\end{center}
\end{figure}

Solar polar field was first proposed to be solar cycle precursor by
Schatten et al (1978), who suggested the correlation between the
polar field near the solar minimum and the subsequent cycle
strength. WSO has the continuous solar polar field observations
since 1976 shown in Figure \ref{fig5} (red curve) overplotted with
the sunspot number \footnote{http://wso.stanford.edu/Polar.html}.
The left panel of Figure \ref{fig6} gives the relation between the
directly observed maximum polar field of cycle $n$ and the
subsequent cycle $n+1$ strength. Cycle 24 is included in using the
predicted strength by Jiang et al.(2007). These 4 points distribute
closely along the straight line which hints the good correlation
between the polar field and the cycle strength.

However the 4 cycle data are less convincing to demonstrate the
correlation. We need longer time series data. Surface flux transport
(SFT) model is an effective way to reconstruct solar polar magnetic
field with the input of sunspot group data (Wang et al. 1989;
Schrijver et al. 2002; Mackay et al. 2002; Baumann et al. 2004).
CJSS10 used the SFT model to reconstruct the surface field and open
flux for the period 1913-1986. The observed sunspot longitudes,
latitudes, areas and cycle-averaged tilt angles were used to create
the source term. The results of that model compare well with the
open flux derived from geomagnetic indices (Lockwood 2003) and the
reversal times of polar fields (Makarov et al. 2003). Right panel of
Figure \ref{fig6} shows the correlation between the reconstructed
polar field maximum during the end of the cycles 15-21 and the
subsequent cycle strength. They show the strong correlation with the
correlation coefficient 0.85. Please note that different definitions
of the polar field were used between the observation (left panel,
line-of-sight component and averaged over 35$^\circ$ wide polar
caps) and the reconstruction (right panel, radial component and
averaged over the 15$^\circ$ wide polar caps). Jiang et al. (2011c)
show that they are consistent when the same definitions are used.
The good correlation provides the evidence for the polar field to be
the solar cycle precursor.

The polar field is the polar component of the solar poloidal field
which is one part of the dynamo loop since the dynamo model is
regarded as the oscillation between the poloidal field and the
toroidal field (Parker, 1955). The toroidal field usually is
corresponding to the cycle strength. The polar field and the cycle
strength are connected by the solar dynamo process. Their
correlation given by the direct or the indirect observations hence
would provide constraints on the solar dynamo.

There are other solar cycle precursors, for example, the length of
solar minimum, the length of solar cycle overlap, the solar activity
before the solar minimum and so on, whose predictive skill may be
explained by the reasons given by Cameron \& Sch\"{u}ssler (2007).
The stronger cycles rise faster toward sunspot maximum. It is the
Waldmeier effect. Moreover, the high-latitude spots of the new cycle
already appear when the old cycle is still in progress in low
latitudes. It is the cycle overlapping. The two properties of the
sunspot record were used to explain the precursor skill. A stronger
follower cycle with a shorter rise time leads to an earlier minimum
and a higher predictor than a weaker subsequent cycle with a longer
rise time. Thus the minimum epoch and height depend on the strength
of the following cycle. Hence no physical connection between the
surface manifestations of subsequent activity cycles is required in
this explanation.

\section{Dynamo mechanisms constrained by the polar field precursor}
\subsection{Possible reasons responsible for the polar field precursor}

Section 2 has shown that the aa-index precursor is consistent with
the polar field precursor since the aa-index at the solar minimum
roughly reflects the strength of polar field maximum by the
connection of the axial dipole component of the open flux. Their
prediction skill provides the constraint on the solar dynamo models.
Let's see the constraints that the precursor skill can give. The
following discussion is based on the Babcock-Leighton (BL)-type flux
transport (FT) dynamo. See Choudhuri (2013) in this proceeding for
more details about the BL-type FT dynamo.

\begin{figure}[htb]
% \vspace*{-1.0 cm}
\begin{center}
\includegraphics[width=3.in]{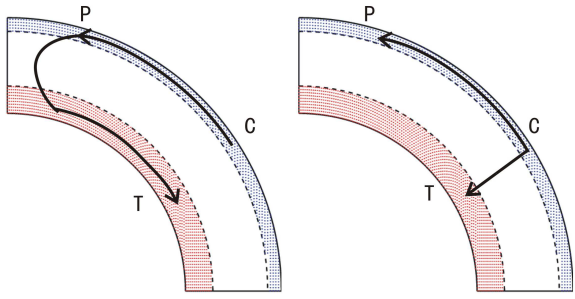}
% \vspace*{-0.5 cm}
 \caption{Cartoons to illustrate the possibilities causing the correlation between
 the polar field (denoted as P) and the next cycle strength (denoted as T).
 The blue and red region show the location of the poloidal and the toroidal
 field generation, respectively. Left panel: the poloidal field generated at the middle latitudes
 of the surface (denoted as C) is first transported to the pole and later to
 the tachocline. Right panel: the poloidal field at C is poleward transported to the pole
 and inward transported to the tachocline simultaneously. See also Jiang et al (2007).}
   \label{fig7}
\end{center}
\end{figure}

Although the correlation between the polar field maximum and the
subsequent cycle strength was proposed by Schatten et al. (1978)
based on the BL-type FT dynamo, here we would like to stress that
not all BL-type FT dynamo can give the correlation. In the BL-type
dynamo, poloidal field is generated at surface layer due to the
decay of the tilt sunspot groups. Toroidal field corresponding to
the subsequent cycle strength locates in the tachocline at the base
of the convection zone. It means that the toroidal and the poloidal
field are spatially separated. There are two possibilities
illustrated by Figure \ref{fig7} to generate the correlation between
two spatial separated parameters (see also Jiang et al, 2007).

During a maximum, the poloidal field created by the BL process is
primarily in mid-latitudes of the surface layer denoted as
\textbf{C} in Figure \ref{fig7}. Left panel shows one possibility.
The poloidal field is transported to the polar region \textbf{P} to
produce the polar field which reaches the maximum near the solar
minimum. Later it is then advected downward to the tachocline, where
it is stretched by the differential rotation to create the toroidal
field. The correlation between the polar field at the minimum and
the strength of the next maximum can be explained if the polar field
can be brought to the mid-latitude tachocline (denoted by
\textbf{T}) in order of 5 yrs, which serves as the constraint of the
dynamo model. The right panel shows the other possibility. When the
poloidal field at \textbf{C} produced during a maximum is swept away
from \textbf{C} polewards to \textbf{P} to form the polar field, it
is simultaneously inward transported to the tachocline at
\textbf{T}. The same cause, polodial field at \textbf{C} causes the
correlation between polar field maximum and the strength of the next
maximum. The constraint in this possibility is that about 5 yrs is
required for the flux at \textbf{C} to transport to \textbf{P} and
T, respectively.

\subsection{Three classes of BL-type FT dynamo models and the comparisons with the precursor constraint}
Meridional flow, turbulent diffusivity and convective pumping are
three potential mechanisms which are responsible for the transport
of the poloidal field from the solar surface to the tachocline. The
convective pumping was once ignored by the past dynamo studies.
Yeates et al. (2008) distinguished the previous BL-type FT dynamo as
the advection (meridional flow) and the diffusion dominated models.
Cameron et al. (2012) indicated that the downward pumping of the
magnetic flux has a significant effect in slowing the diffusive
transport of magnetic flux through the solar surface. Provided the
pumping was strong enough, the FT dynamo using a vertical boundary
condition matches the SFT model. This study is supposed to be the
prelude to initiate the pumping dominated BL-type FT dynamo models.

In the following we list some typical models of each class and
analyze that whether each class is consistent with the precursor
constraints. The meridional flow is a basic ingredient in all the
BL-type FT dynamo models. Helioseismology has few knowledge on the
equatorwards meridional flow in the solar interior. Hence different
profiles were adopted in the dynamo models. In spite of the
differences, there is no much difference for the poloidal field to
be transported to the bottom under the sole advection of the
meridional flow. It is about 2 cycles for all the models.

\subsubsection{Meridional flow dominated class}

The dynamo models aiming for the solar cycle prediction used in
Dikpati et al. (2006a, 2006b) belong to the meridional flow
dominated class. Turbulent diffusivity in their model is in the
range of $3\times 10^{10}$ -- $2\times10^{11} cm^2 s^{-1}$. It takes
more than 5 solar cycles for the poloidal field at the surface to
diffuse all the way to the bottom. No convective pumping was
considered. Hence the transport mechanisms in their models are the
meridional flow dominated.

Due to the advection of the meridional flow, the poloidal field is
transported to the poles to form the polar field and later to the
tachocline where it is stretched to form the toroidal field. This is
consistent with the fist possibility mentioned in the last section
(left panel of Figure \ref{fig7}). However, it takes 17-21 years for
the meridional flow to take the polar field to the tacholine. Hence
it causes the correlation between polar field and the strength of
the cycle after the next. Obviously, this is not consistent with the
polar field precursor constraint.

\subsubsection{Turbulent diffusion dominated class}
Another group of dynamo-based solar cycle prediction models given by
Choudhuri et al. (2007) and Jiang et al. (2007) (see also Chatterjee
et al., 2004) are turbulent diffusion dominated. The turbulent
diffusivity is in order of $10^{12} cm^2 s^{-1}$. About 5 yrs are
required for the poloidal field at the surface to be diffused to the
bottom. Hence the models belong to the turbulent diffusion dominated
class.

When the poloidal field is advected to the pole under the effect of
the meridional flow and the turbulent diffusion to form the polar
field, the poloidal field is diffused to the tachocline
simultaneously mainly due to the high diffusivity. This is
consistent with the second possibility in the last section (right
panel of Figure \ref{fig7}). The transport time from the surface
middle latitudes to the pole and to the tachocline are both about 5
yrs. Thus in the model, the polar field at the solar minimum has a
strong correlation with the subsequent cycle strength. Obviously,
the result shown in this group of models is consistent with the
polar field precursor. The high diffusivity in the convective zone
is proved to be a possibility for the surface poloidal flux to be
transported to the tachocline.

\begin{figure}[htb]
% \vspace*{-1.0 cm}
\begin{center}
\includegraphics[width=.6in]{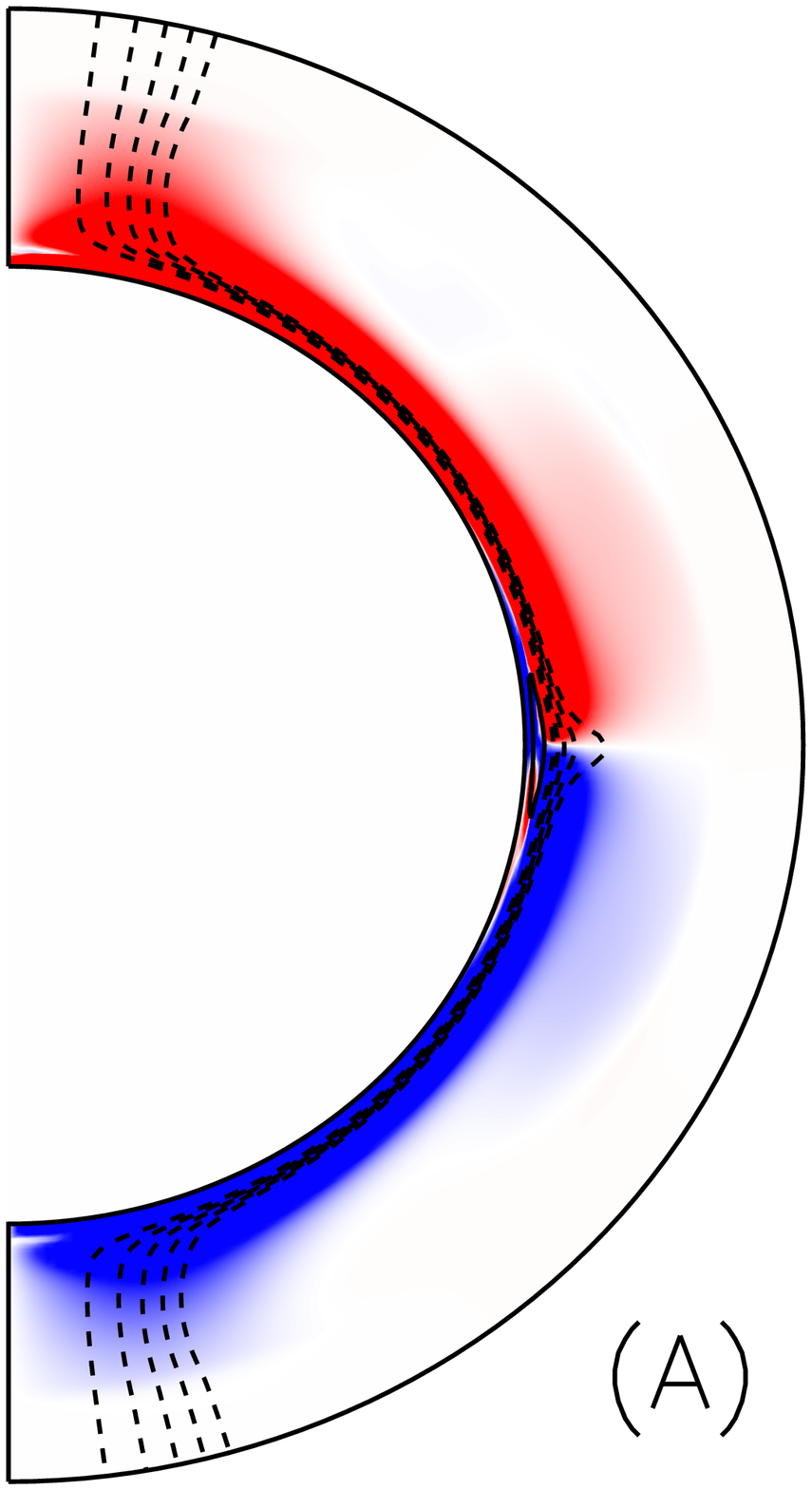}
\includegraphics[width=.6in]{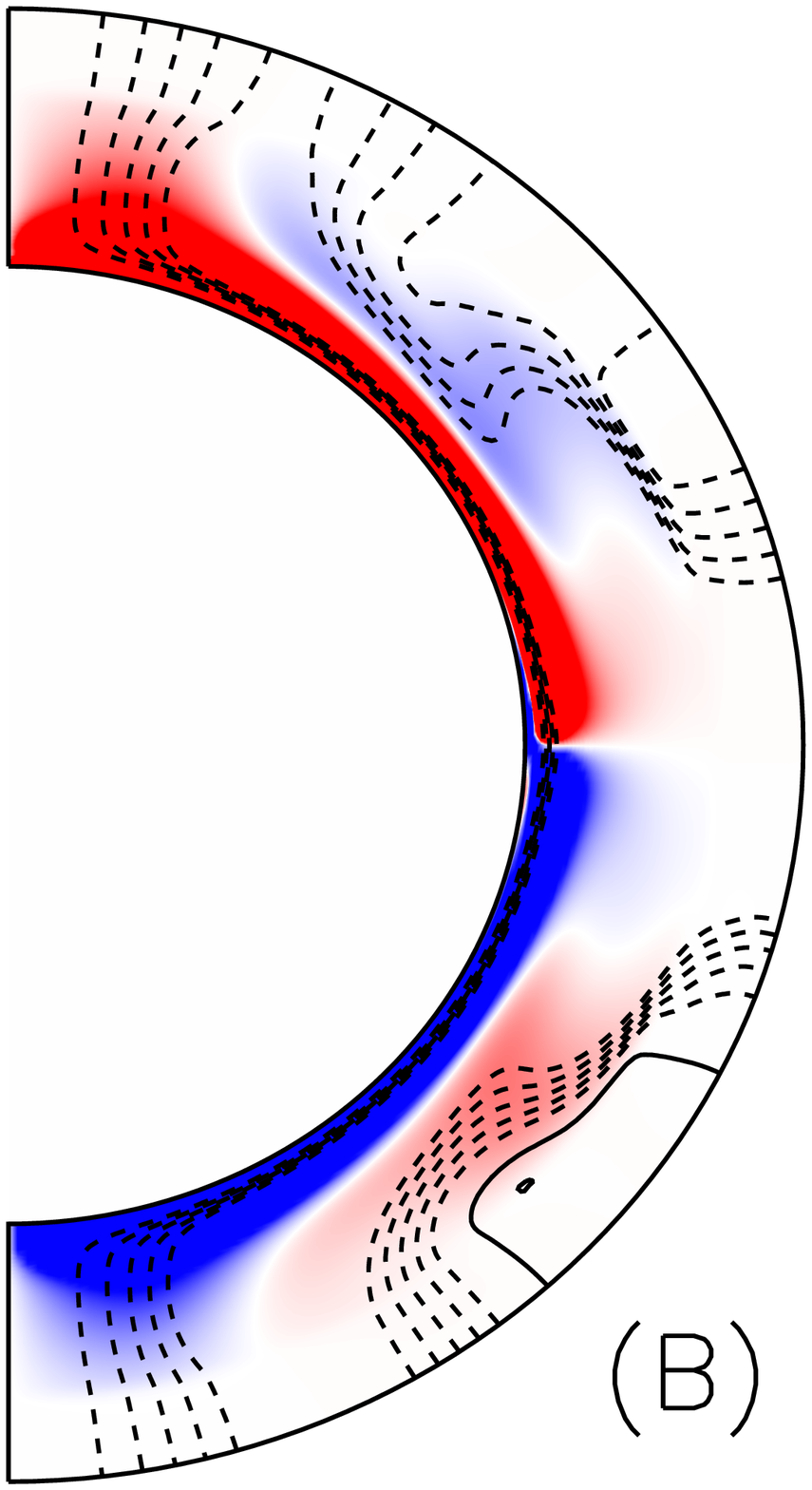}
\includegraphics[width=.6in]{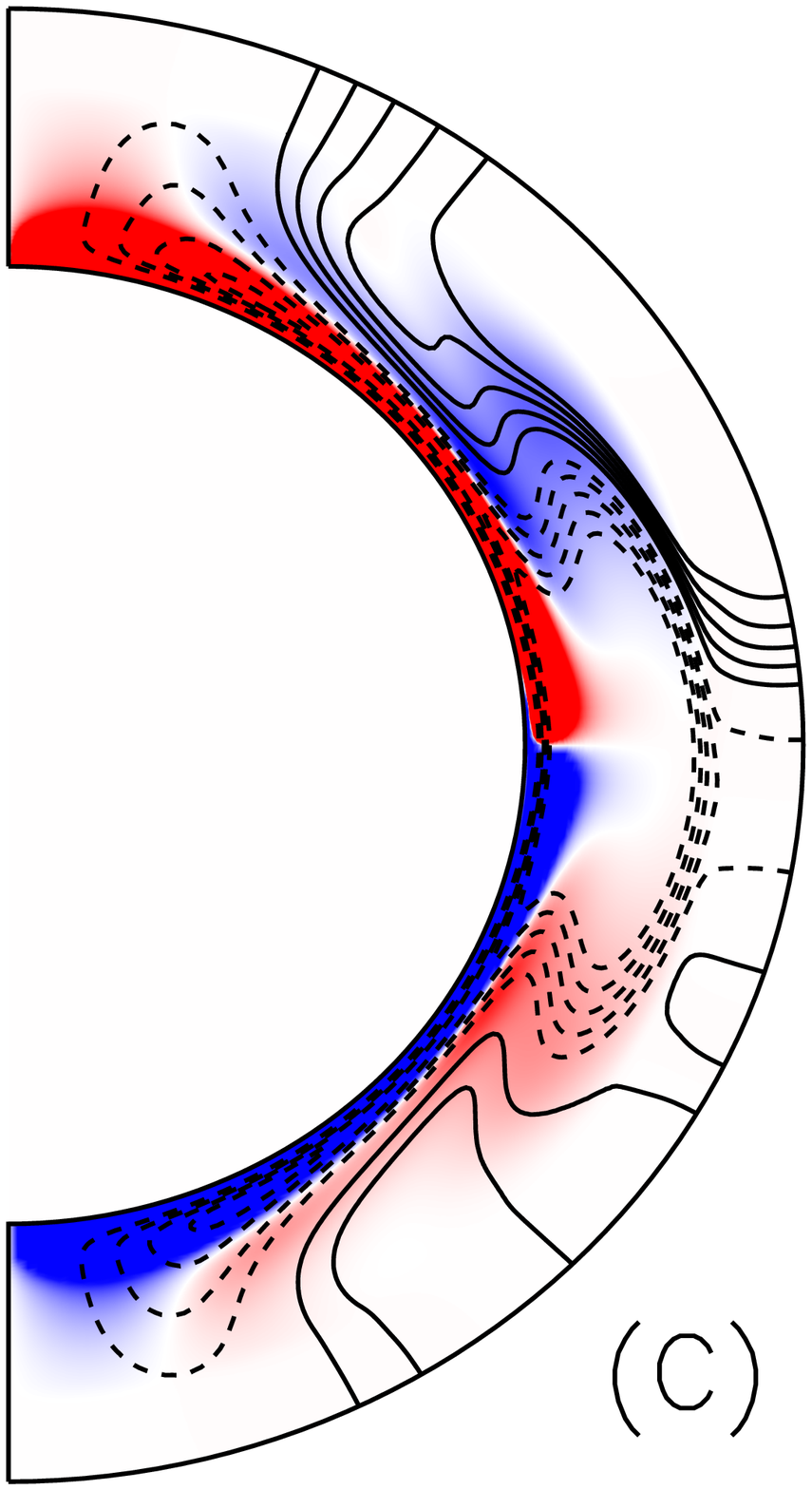}
\includegraphics[width=.6in]{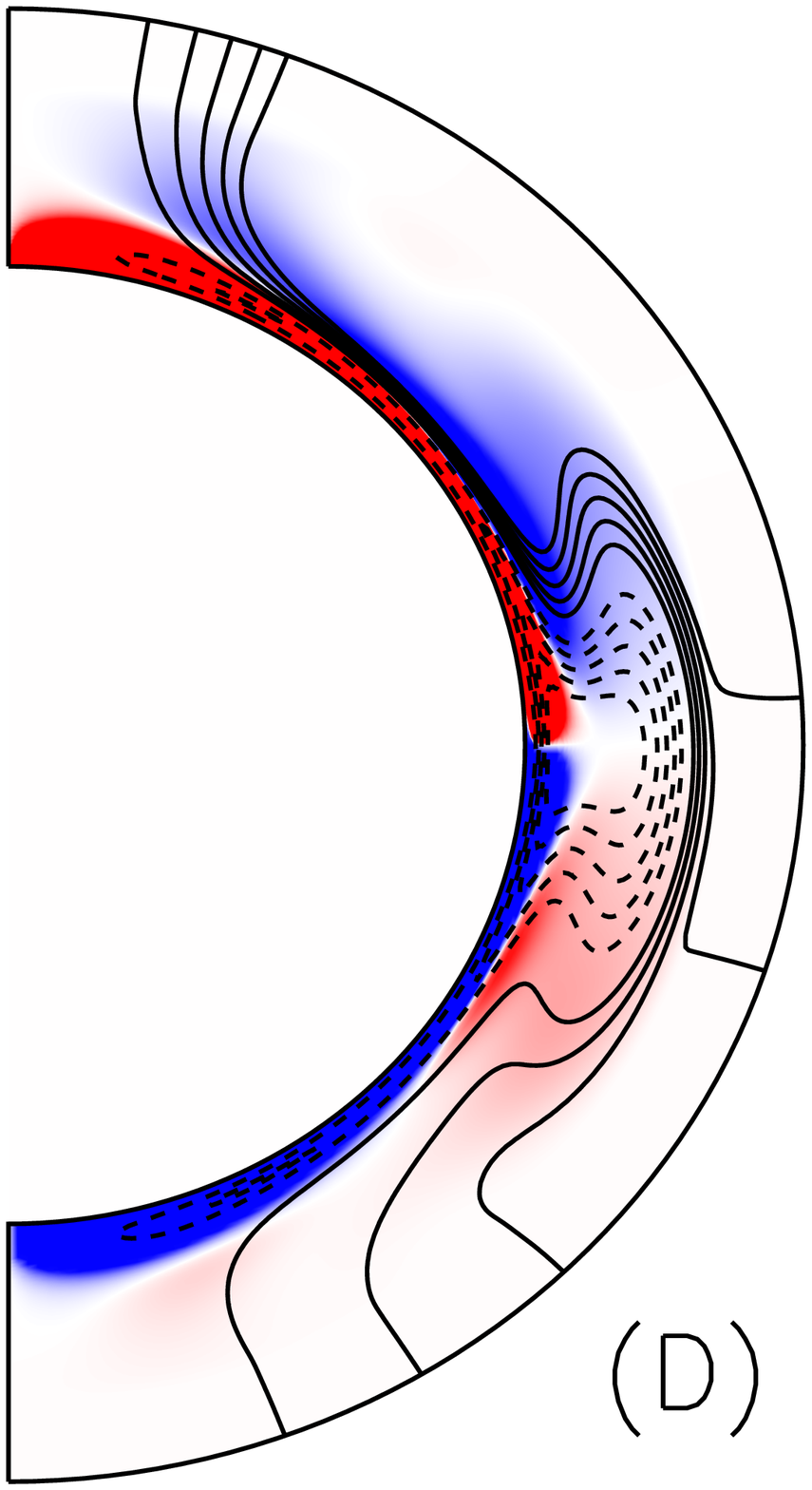}
\includegraphics[width=.6in]{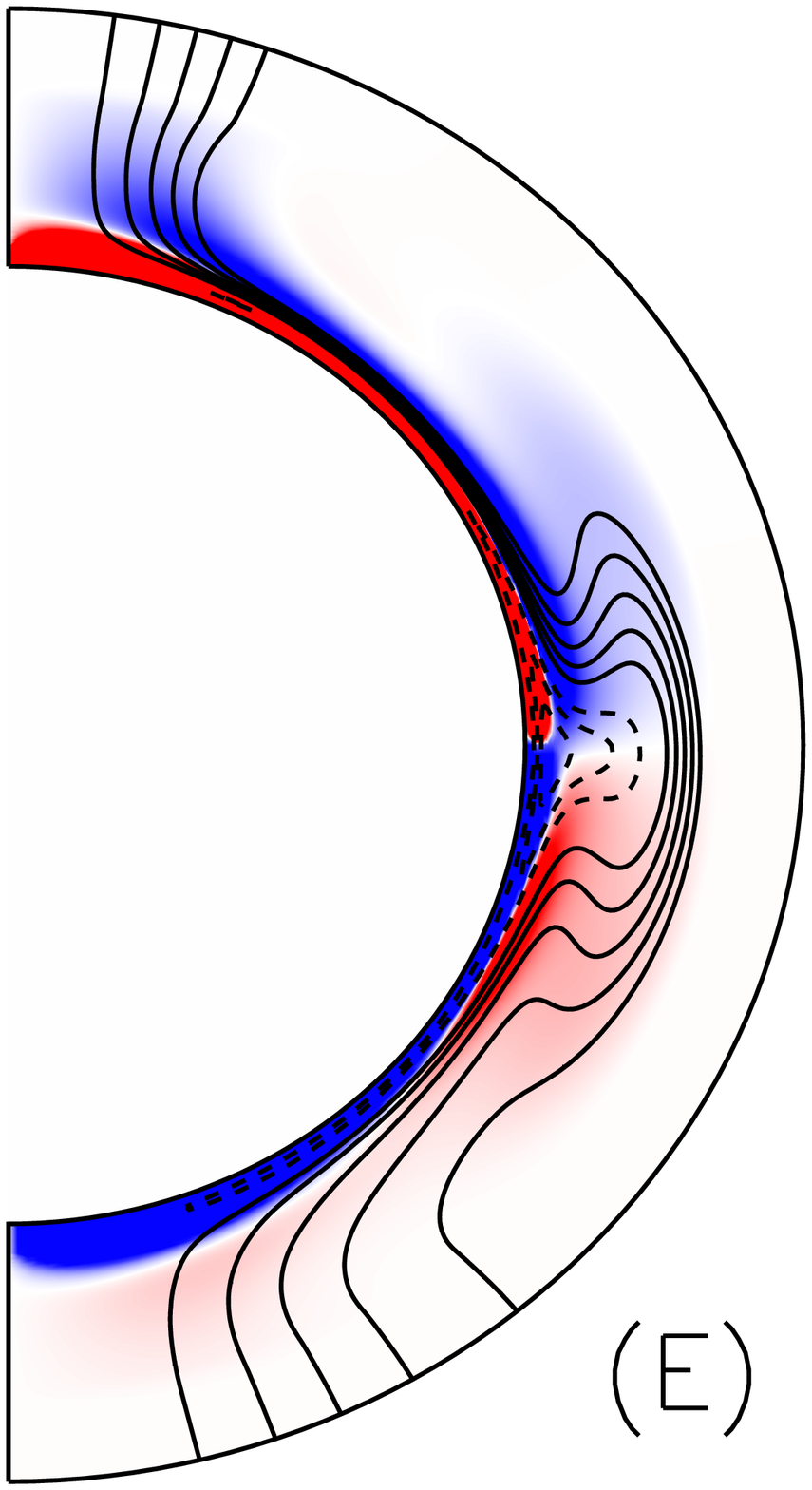}
\includegraphics[width=.6in]{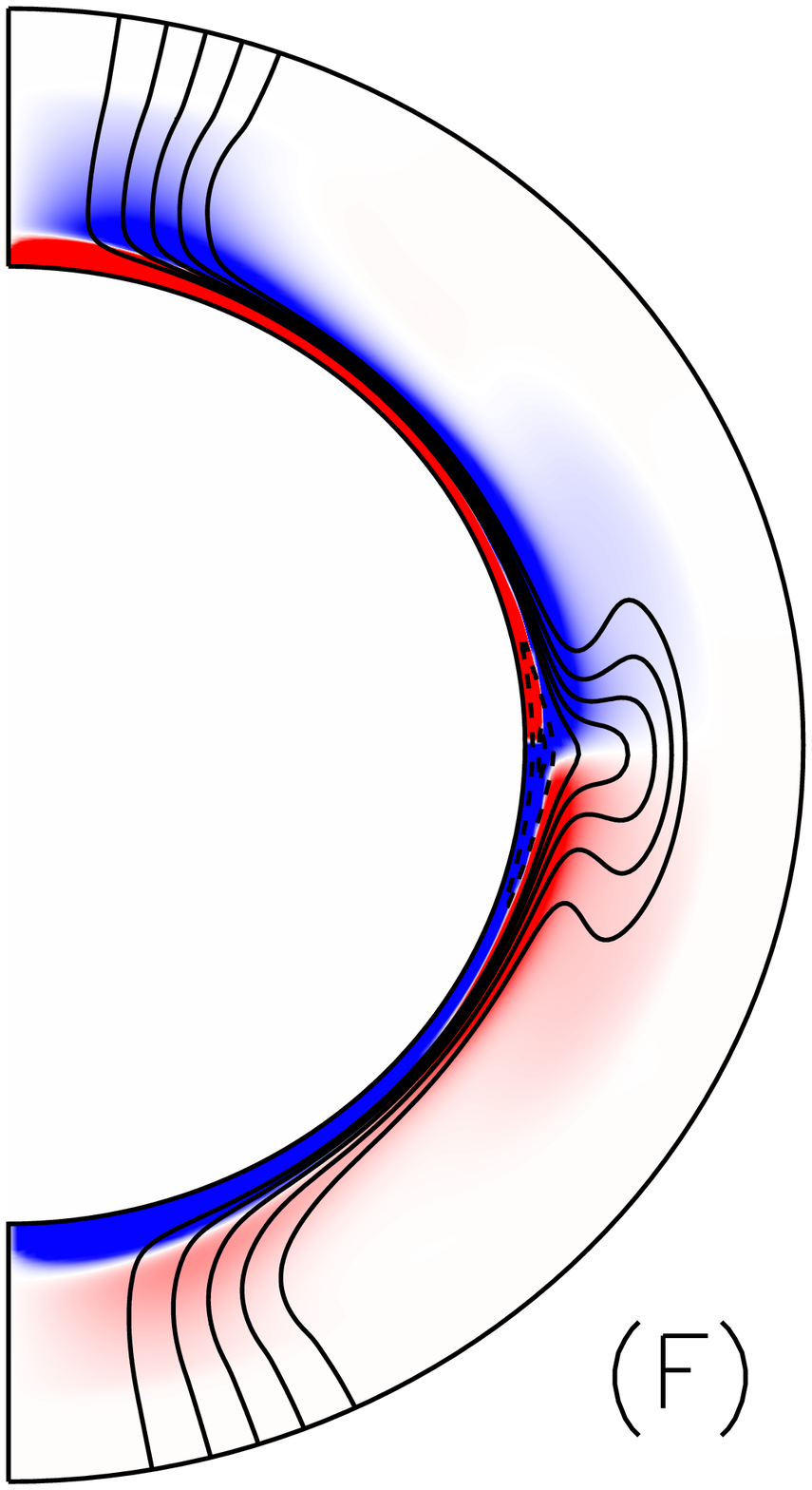}
\includegraphics[width=3.0in]{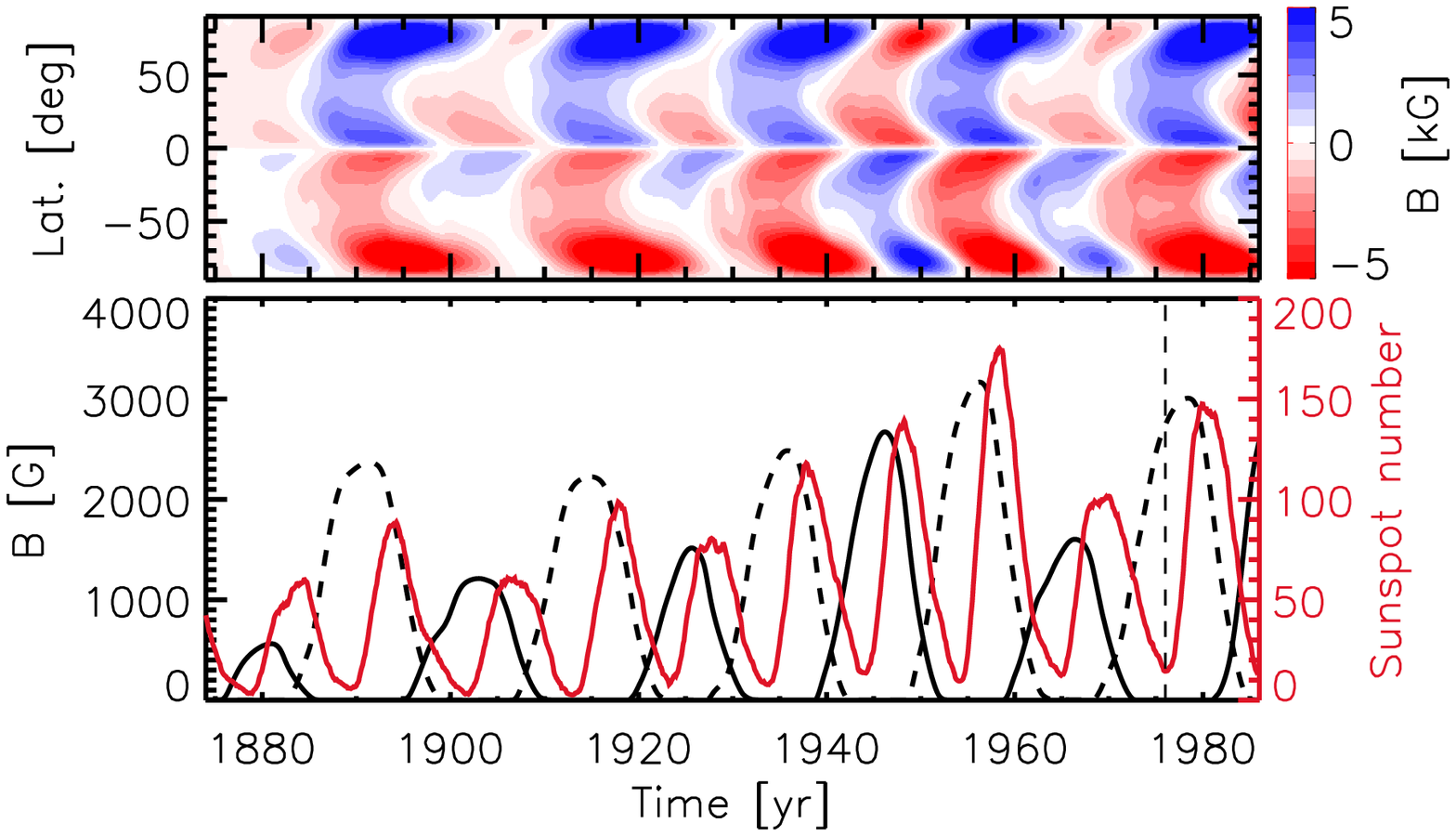}
%\includegraphics[width=1.7in,angle=90]{fig8b_2.ps}
% \vspace*{-0.5 cm}
\caption{Results of the convective pumping dominated model from
Jiang et al. (2013). Top panel: variation of the magnetic field over
solar cycle 19. The red and blue colors correspond to negative and
positive toroidal flux respectively. The lines are field lines based
on the poloidal flux. Middle panel: the evolution as a function of
latitude and time of the toroidal field at $r=0.7R_\odot$. Lower
panel: the average unsigned toroidal flux between $¡À45^{\circ}$
latitude corresponding to odd and even cycles shown using black
dashed and solid curves, respectively.}
   \label{fig8}
\end{center}
\end{figure}

\subsubsection{Convective pumping dominated class}

Recently people become aware of the effects of convective pumping in
the BL-type FT dynamo. Guerrero \& de Gouveia Dal Pino (2008),
Cameron et al. (2012), Karak \& Nandy (2012) and Jiang et al. (2013)
have shown the importance of the convective pumping in their models.
In the following the model given by Jiang et al. (2013) is presented
to show how pumping works in the FT dynamo models.

The magnetic pumping extends to the base of the convection zone with
the strength 2ms$^{-1}$ in the convection zone. Near the surface, it
reaches 20ms$^{-1}$ which is consistent under the constraint by
Cameron et al. (2012) to prevent the diffusive transport of magnetic
flux through the solar surface. Less than 5 yr is required for the
poloidal field at the surface to be transported to the bottom under
the sole effect of turbulent pumping. Hence the transport mechanism
in the model is pumping dominated. The turbulent diffusivity is
$2\times10^{11} cm^2 s^{-1}$.

The poloidal source was based as closely as possible on the
observations. The time evolution of the magnetic field through solar
cycle 19 is shown in the upper panel of Figure \ref{fig8}. The
transport of poloidal flux from the surface to the tachocline occurs
over the same time span as the flux is transported to the poles.
Throughout the cycle both the poloidal and toroidal fluxes have a
simple form, i.e. no remnants from previous cycles. In middle panel
of Figure \ref{fig8} we see equatorward propagating toroidal flux at
low latitudes. The maxima of the modeled activity levels at $r =
0.7R_\odot$ and over latitudes $-45^\circ < \lambda < 45^\circ$ are
clearly related to the amplitude of the observed cycles (lower
panel). The correlation coefficient for cycles 15 to 21 is 0.93. The
surface evolution of the field is similar to that given by the
surface flux transport model. The correlation between the polar
fields and the strength of the next cycle is 0.85.

Hence the results from the model with the pumping dominated
transport mechanism is consistent with the constraint by the polar
field precursor as well. The turbulent pumping provides the other
possibility to dominate the transport of the surface poloidal flux
to the bottom.

\section{Solar polar (surface poloidal) field generation}
\subsection{Overview of the polar field generation}
Sections 2 and 3 have shown the importance of the solar surface
poloidal field which could connect the observations with the BL-type
FT dynamo models. As the polar component of the poloidal field,
solar polar field is a direct observational quality (the consistent
measurement started from 1976 as mentioned before). Now we analysis
the polar field generation which is related to the solar poloidal
field generation.

\begin{figure}[htb]
% \vspace*{-1.0 cm}
\begin{center}
\includegraphics[width=2.in]{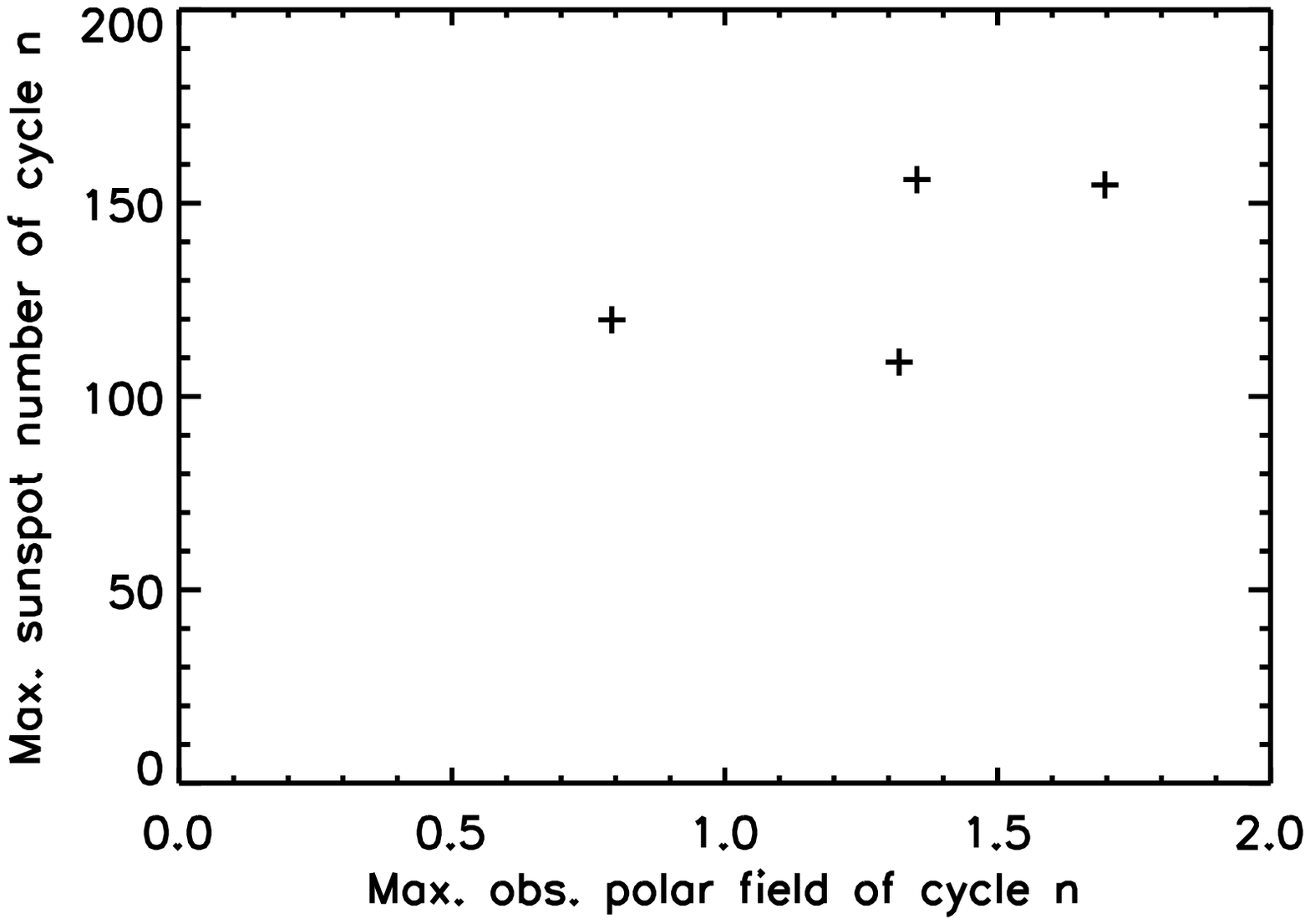}
\includegraphics[width=1.7in]{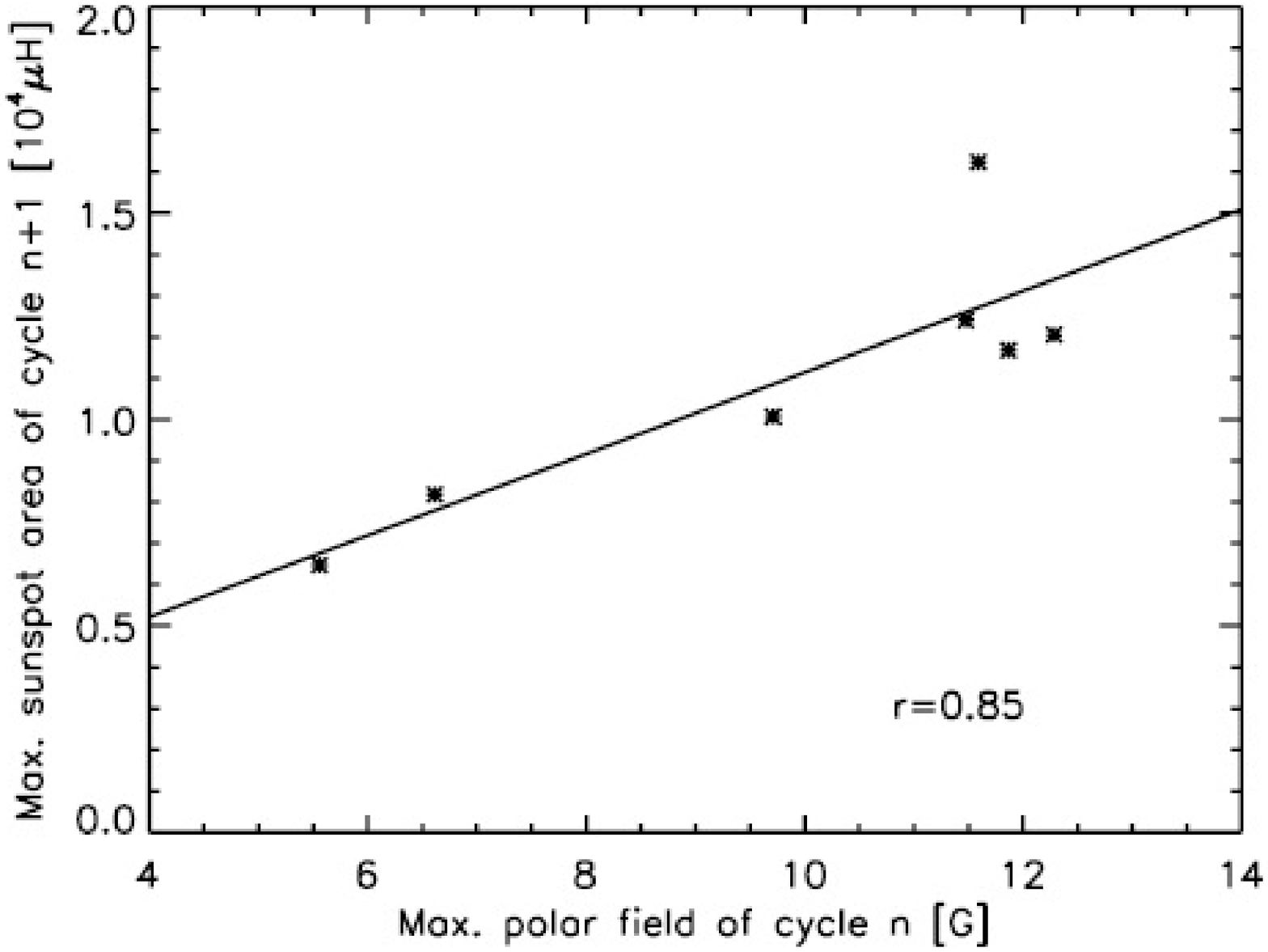}
% \vspace*{-0.5 cm}
 \caption{Same as Figure \ref{fig6}, but for the correlations between the polar field
 and the same cycle strength, which shows no correlation.}
   \label{fig9}
\end{center}
\end{figure}

Left panel of Figure \ref{fig9} is the maximum values of the
observed polar field (from WSO) at the end of each solar cycle verse
the maximum sunspot number of the same cycle during cycles 20-23.
Right panel is the reconstructed solar polar field given by Cameron
et al. (2010) verse the maximum sunspot areas of the same cycle
during cycles 15-20. Being contrary to Figure \ref{fig6}, we may see
that there is no correlation between the polar field and the same
cycle strength.

The evolution of the solar polar field may be described as follows.
The eruption of bipolar sunspot groups with their leading polarities
systematically equatorward of their following polarities establishes
an overall separation of polarities in latitudes. Differential
rotation and surface diffusion separate the two polarities further.
The flux of leading polarities decays around the equator due to the
effect of diffusion. Meridional flow and the turbulent diffusion
transport the net flux of the following polarities in each
hemisphere to the poles to reverse the polar field of old cycle and
build up the polar field of new cycle. In summary, the tilt sunspot
group emergence provides the solar polar field source. The advection
affects the polar field evolution. The nonlinearities and the
randomness are involved in the sunspot group emergence, and the
meridional flow possibly varies over solar cycles. Eventually, the
polar field strength is dominated by the cross-equator net flux
during the evolution of the solar cycle.

\subsection{Nonlinearities in the polar field generation}
\begin{figure}[htb]
% \vspace*{-1.0 cm}
\begin{center}
\includegraphics[width=2.0in]{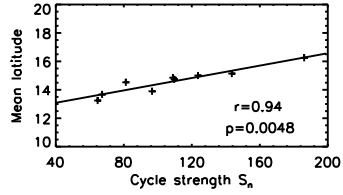}
% \vspace*{-0.5 cm}
 \caption{Correlation between
cycle averaged latitudes of sunspot emergence and the cycle strength
defined by the maximum sunspot number during cycles 12-20 (from
Jiang et al., 2011a).}
   \label{fig10}
\end{center}
\end{figure}

Jiang et al. (2011a) studied the statistical properties of sunspot
group emergence on the cycle phase and strength. One of their
results shown in Figure \ref{fig10} gives the correlation between
cycle averaged latitudes of sunspot emergence and cycle strength
defined by the maximum sunspot number. We see that strong cycles
tend to have higher mean latitudes for sunspot group emergence. High
mean latitude for sunspot group emergence generates less
cross-equator flux, hence a weak polar field. This is one nonlinear
mechanism to modulate the polar field generation for strong cycles.

\begin{figure}[htb]
% \vspace*{-1.0 cm}
\begin{center}
\includegraphics[width=4.0in]{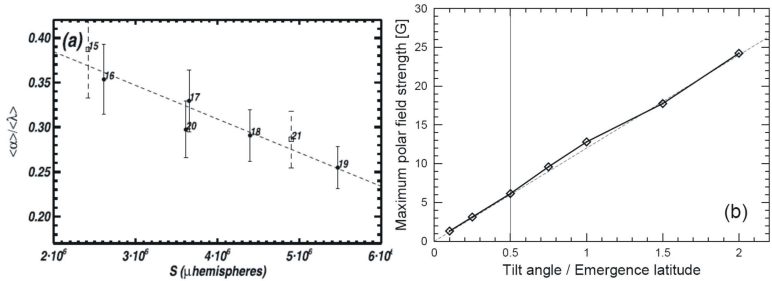}
% \vspace*{-0.5 cm}
 \caption{Nonlinearity in polar field generation caused
 by the tilt angles of sunspot group emergence. Left panel:
 cycle averaged tilt angle normalized by the emergence latitude
vs. strength of the same cycle (from Dasi-Espuig et al., 2010).
Right panel: polar field strength vs. the tilt angles (from Baumann
et al., 2004).}
   \label{fig11}
\end{center}
\end{figure}

The tilt emergence (related to Joy's law) is a distinguished feature
of sunspot groups. Left panel of Figure \ref{fig11} shows the study
from Dasi-Espuig et al. (2010) about the cycle averaged tilt angle
normalized by the emergence latitude verse the strength of the same
cycle. We see that the strong cycle corresponds to weak sunspot
group tilts. Right panel of Figure \ref{fig11} is the result from
Baumann et al. (2004) about the polar field strengths and the
sunspot group tilts which present a good linear relation. For weaker
tilt angles, cross-equatorial cancelation of preceding flux is
decreased leading to the accumulation of less flux at the poles.
Hence the weaker tilt angles of sunspot groups for stronger cycles
is another nonlinearity to modulate the solar polar field
generation.

The features of sunspot group emergence that the cycle averaged tilt
angles are weaker and the mean latitudes are higher for the stronger
cycles provide the nonlinearities to modulate the polar field
generation.

\subsection{Randomness in the polar field generation}
\begin{figure}[htb]
% \vspace*{-1.0 cm}
\begin{center}
\includegraphics[width=5.0in]{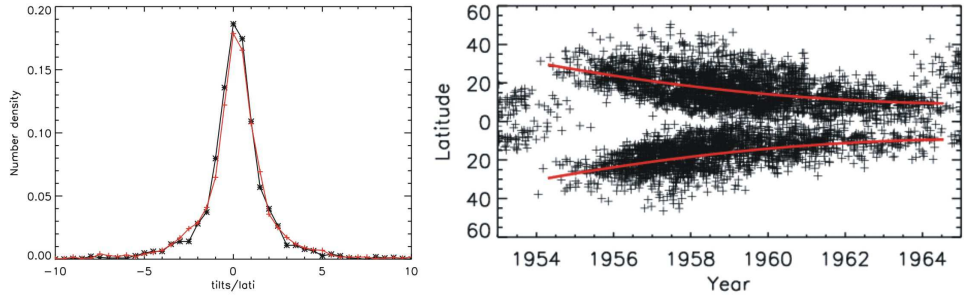}
% \vspace*{-0.5 cm}
 \caption{Randomness of the sunspot group emergence. Left panel: number
density distribution of the sunspot group tilts based on the MWO
(black) and Kodaikanal (red) tilt angle data; Right panel: latitude
distribution during solar cycle 19. Red curve denotes the mean
latitudes.}
   \label{fig12}
\end{center}
\end{figure}

Except the nonlinearities, the polar field generation also involves
in the random processes. The well anti-correlation between the tilt
angle and the cycle strength shown in Figure \ref{fig11} is a cycle
averaged behavior. Left panel of Figure \ref{fig12} shows the number
density distribution of the observed sunspot group tilts based on
the MWO and Kodaikanal tilt angle data. We see the tilt angle value
presents a large scatter probably due to the effect of convective
flows on the flux tube emergence (Longcope \& Fisher, 1996).
Concerning the latitude distribution shown in right panel of Figure
\ref{fig12}, the mean latitude at different phase of solar cycle
denoted by the red curve may be given by a well defined second
polynomial (see Jiang et al., 2011). But the wide latitude
distribution indicates the scattering distribution in the latitudes
at given cycle phases. That how strongly these unpredictable random
processes affect the polar field generation will be studied in our
future work.

\subsection{Fluctuations in the meridional flow}
The parameter studies given by Baumann et al. (2004) indicate that
the polar field as a function of the meridional flow ($v_0$)
initially increases as more following polarity flux is carried to
the poles. As $v_0$ becomes even larger, both magnetic polarities
are carried to the pole by the flow and cancel there. This leads to
a reduction of the polar field.

Jiang et al. (2011) use a well-calibrated surface flux transport
models to study the weak polar field of cycle 23. They find that the
low polar field of cycle 23 could be reproduced by an increase of
the meridional flow by 55\% in cycle 23. Using the flux transport
dynamo model, Karak (2010) and Karak \& Choudhuri (2011) studied the
effects of the meridional flow fluctuations on the solar cycle
irregularities.

\section{Conclusions}
In the paper, I have a review of two major solar cycle precursors,
the geomagnetic variations and the solar polar field. The physical
reasons responsible for their precursor skill are presented. Further
the constraints that the polar field precursor provides to the solar
dynamo model are discussed and the physical mechanisms modulating
the polar field generation are proposed. The main points can be
concluded as follows.

1. The geomagnetic variation (the aa-index) precursor is consistent
with the polar field precursor since the aa-index at the solar
minimum roughly reflects the strength of polar field maximum by the
connection of the axial dipole component of the open flux.

2. The polar field precursor hints that the transport of the
poloidal field generated at the solar surface to the tachocline in
less than one cycle, which provides a constraint on the flux
transport mechanism. The BL-type FT dynamo models with the high
turbulent diffusivity (in order of $10^{12}cm^2s^{-1}$) or the
convective pumping (in order of 2ms$^{-1}$) satisfy the constraint.

3. Solar cycle strength is predictable at the end of the preceding
cycle since the generation of the toroidal field from the poloidal
field is regarded as a deterministic process. However, the
generation of the poloidal field is a complex process. The
nonlinearities and the random processes are involved in the sunspot
emergence. The flux transport also probably involves the
fluctuations. These factors may be the cause of the solar cycle
irregularities. Other unknown nonlinearities and the random
processes are also possible involved.

\acknowledgments{The author acknowledges the financial support from
the National Natural Science Foundations of China (11173033,
11178005, 11125314) and the Knowledge Innovation Program of CAS
(KJCX2-EW-T07).}

\end{document}